\documentclass{article}
\usepackage{latexsym}

\font\sf=cmss10 at 10pt 

\font\bb=msbm10 at 10pt

\font\rms=cmr10 at 8pt
\font\rmss=cmr10 at 7pt

\font\cal=cmsy10 at 9pt
\font\cals=cmsy10 at 8pt

\def\0#1{\mbox{\rm#1}}
\def\1#1{\mbox{\bb#1}}
\def\2#1{\mbox{\bf#1}}
\def\3#1{\mbox{\cal#1}}
\def\4#1{\mbox{\cals#1}}
\def\5#1{\mbox{\sf#1}} 
\def\6#1{\mbox{\rms#1}}
\def\7#1{\mbox{\bbs#1}}
\def\8#1{{\tilde #1}}
\def\9#1{{\breve #1}}

\def\BEn{\begin{enumerate}}
\def\EEn{\end{enumerate}}
\def\BEq{\begin{equation}}
\def\EEq{\end{equation}}
\def\BEqA{\begin{eqnarray}}
\def\EEqA{\end{eqnarray}}

\def\wt{\widetilde}
\def\wh{\widehat}

\def\al{\alpha}

\def\be{\beta}

\def\de{\delta}

\def\si{\sigma}

\def\tav{\hbox{
\kern-1.0pt
\rule[0pt]{1.3pt}{.8pt}{\kern-3.6pt}
\rule[0pt]{.4pt}{6pt}{\kern-3.0pt}
\rule[4.5pt]{3.0pt}{.8pt}{\kern-3.3pt}
\rule[0pt]{.4pt}{5pt}{\kern-1pt}
}}

\def\dag{\dagger}
\def\dar{\downarrow}
\def\lar{\leftarrow}
\def\uar{\uparrow}
\def\adj{{^{\dag}}}

\def\x{\times}

\def\Bar{\kern5pt{\rule[-2.5pt]{.6pt}{9.5pt}}\kern5pt}
\def\bra{\langle}

\def\ket{\rangle}

\def\II{|\kern-1pt |}

\def\ox{\otimes}

\def\pd{\partial}

\def\lmult{{\lfloor\kern-5pt\lfloor}}
\def\rmult{{\rfloor\kern-5pt\rfloor}}

\def\Cliff{\mathop{{\rm Cliff}}\nolimits} 

\def\Dim{\mathop{\hbox{\rm Dim}}\nolimits}

\def\End{\mathop{\rm Endo}\nolimits} 

\def\SO{\mathop{\mbox{\rm SO}}\nolimits}

\def\Sq{\mathop{\hbox{\rm Sq}}\nolimits}  

\begin{document}

\title
{
CLIFFORD ALGEBRA \\AS QUANTUM LANGUAGE\footnote
{
Some of these results were presented at the
American Physical Society Centennial Meeting,
Atlanta, March 25, 1999
[Marks, Galiautdinov, Shiri, Finkelstein,
Baugh, Kallfelz, and Tang (1999)].
}
}

\author{James Baugh \quad
David Ritz Finkelstein
\quad Andrei Galiautdinov\\
{\normalsize School of Physics,
Georgia Institute of Technology}
\and
Heinrich Saller\\
{\normalsize Heisenberg Institute of Theoretical Physics,
Max Planck Society}
}

\date{\today}
\maketitle

\abstract
{
We suggest Clifford algebra as a useful simplifying
language for present quantum dynamics.
Clifford algebras arise from representations of the permutation groups
as they arise from representations
of the rotation groups.
Aggregates using such representations for their permutations
obey {\em Clifford statistics}\/.
The vectors
supporting the Clifford algebras of permutations and rotations
are plexors and spinors respectively.
Physical spinors may actually be plexors
describing quantum ensembles, not simple individuals.
We use Clifford statistics
to define quantum fields on a quantum space-time,
and to formulate a
quantum dynamics-field-space-time unity
that evades the compactification problem.
The quantum bits of history regarded as a quantum computation
seem to obey a Clifford statistics.
}

\newpage

\pagestyle{myheadings}
\markright{J. Baugh, D. R. Finkelstein, A. Galiautdinov, H.
Saller, {\bf Clifford algebra as quantum language} }

\section{Spinors and plexors}
\label{sec:SAP}
Pauli represented\footnote
{
In what follows,
all representations and their homomorphisms
are projective,
and may be double-valued,
unless they are stated to be linear or vector representations,
which are single-valued.
$\21, \22,
\23, \dots$ represent real quadratic spaces
of dimension $1, 2, 3, \dots$ and of signature specified in context.
}
   electron rotations in $\SO(3)$
with elements of the Clifford algebra of a $\23$,
and Dirac represented Lorentz transformation in $\SO(1,3)$
with elements of the Clifford algebra of a $\24$.
It is unlikely that they knew that
Wiman (1898) and Schur (1911) had represented permutations
in the permutation group\footnote
{
Also called ``the symmetric group,''
no doubt because its elements are not symmetric.
}
$\0S_{N}$ with elements of the Clifford algebra of $(N-1)\21$.
A vector space (or module) on which these Clifford algebras
are faithfully represented as endomorphism algebra (or ring)
are called spinors for the orthogonal groups and {\em
plexors} for the permutation groups. Plexors and spinors are
isomorphic mathematical objects
of different physical meaning.
Spinors arise in 1-body quantum physics,
plexors in $N$-body quantum physics with $N>3$.

In the first years of quantum theory,
physicists overlooked spinors
because they
do not occur in the tensor product
of  vectors.
We then proceeded to overlook plexors until Nayak \& Wilczek
(1996) for much the
same reason.

\subsection{Clifford representation of the permutation groups}
We write the free Clifford algebra over a quadratic space $V$
as
\BEq
\Cliff(V)={\22}^V
\EEq
and the spinor space $S$ of $V$ defined by
the isomorphism ${\22}^V\cong S\ox S\adj$,
as
\BEq
S=\sqrt{\22^V}=\sqrt{\22}^V\/.
\EEq
We construct a reducible Clifford representation $R$ of the permutation
group $\0S_N$
by associating the $a$-th individual of the $N$-ad being permuted
with a first-degree Clifford unit
   $i_a\in \22^{N\bf 1}$,
for $a= 1,\dots , N$,
obeying
Clifford relations
\BEq
\label{eq:CLIFF}
i_a i_b + i_b i_a =2 \de_{ab}.
\EEq
For quantum applications we define an adjoint $\dag$ on $\22^V$ by
\BEq
i_a\adj=i_a.
\EEq
Then $R$ represents each swap $(ab)\in \0S_N$ (for $a < b$) by
the Clifford difference $R(ab)=i_b-i_a$:
\BEq
\label{eq:DIFF}
R: \0S_N\to  \22^{N\bf 1}, (ab)\mapsto i_a - i_b\/.
\EEq

It is straightforward to see that $R$ is a (projective!)
representation of $\0S_{N}$.
The $i_a$ make up an orthonormal basis
for the first-degree subspace  $\3V=\Cliff_1(N,0)\subset\Cliff(N,0)$.
Let us call $R$ the {\em orthonormal} Clifford representation of $\0S_N$.

$R'$, a useful variant of $R$, replaces (\ref{eq:CLIFF}) by
\BEq
\label{eq:CLIFF2}
i_a i_b + i_b i_a =2 t_{ab},
\EEq
where
\BEq
\label{eq:SIMPLEX}
t_{ab}={N+1\over N-1}\delta_{ab}-{2\over N-1}\/.
\EEq
The form $t_{ab}$ is the inner product
between two unit vectors $i_a, \, i_b$ from the center of a
regular
$N-1$-simplex
$\si^{N-1}\subset (N-1) \21$ to its $N$ vertices $i_1, \dots,
i_N$. The $i_a$ of (\ref{eq:CLIFF2}) can be identified with the
$N$ simplex vertices permuted.
We therefore call $R'$ the {\em simplicial} representation of $\0S_{N}$.
It is isomorphic to
a representation given less symmetrically by Schur (1911).
The $R'$ image of $\0S_N$ is an irreducible group of Clifford-algebra elements
$\wt{\0S}(N)$ that is a {\em (universal) covering group}\footnote{
Schur (1911) would call this a ``representation group'' of
$\0S_N$,
but today this seems apt to be confused with a group representation.}
   $\wt{\0S}_N$ of
the discrete group $\0S_N$.

The concept of rotation presupposes physical concepts
of angle and length.
The concept of permutation does not.
It presupposes the more primal concept of identity.
Therefore
permutation groups can enter theoretical physics at finer levels of resolution
and higher energies than
rotation groups.
Plexors may be more basic than spinors.

Representations of permutation groups, and hence plexors, enter quantum physics
by two specially deep routes, statistics and dynamics.

\section{Clifford statistics}

A {\em statistics}
describes how we compose individual
quantum elements into
an aggregate quantum system.
Such a transition from the individual
to the collective
is sometimes called ``second quantization,''
merely because it introduces new operators,
but
is better called quantification.
Quantification has little to do with quantization
and is thousands of years older.

A quantification or statistics is often defined by
giving a representation of each
permutation group $\0S_N$.
A {\em Clifford statistics} represents
permutations doubly
by Clifford algebra elements of degree 1 [Finkelstein \&
Galiautdinov (2000)]. A {\em cliffordon} is a quantum whose
aggregates have Clifford statistics. A {\em squad} is a quantum
assembly of cliffordons.

The operators representing operations on a squad of cliffordons,
including the observables of the squad,
form the Clifford algebra of the one-cliffordon mode space.
Modes of the squad
are plexors of its Clifford
algebra,
representable by elements of a minimal left ideal
of the algebra.

A statistics is {\em abelian}
(or central, or scalar, respectively)
as its representation of the
permutation group is.
The Bose and Fermi statistics are abelian,
and the Maxwell, Clifford and braid statistics are non-abelian.
Read \& Moore (1992)
suggested non-abelian statistics for quasi-particles
of the fractional quantum Hall effect.
Nayak \& Wilczek (1996) recognized this as a Clifford statistics
in the first application of the Schur (1911) theory to physics.
While anyonic and braid statistics are confined to two dimensions,
Clifford statistics
works in any dimensionality.

\subsection{Spinors describe aggregates}
When a spinor of $\SO_{N}$ describes
one quantum individual with $N$ possible modes,
a plexor of $\0S_N$ describes a complex of at least $N$ isomorphic
individuals.
Thus our belated encounter with Schur (1911)
disabuses us of a
long-held notion
that
spinors represent simpler entities
than vectors,
a notion that has blocked important research directions.
Spinors represent aggregates.

This deserves to be said three times, differently:

\subsection{Spinors describe aggregates}

That spinors describe aggregates in quantum theory
was already implicit in Cartan's theory of spinors.
Cartan starts from a complex vector space $\3V$
with a symmetric quadratic form $g$.
The complex space $\3V$ can be decomposed
(in many ways)
into two maximal null subspaces $\3V_{\pm}$
as $\3V=\3V_+\oplus \3V_-$.
(A {\em null space} is one composed entirely of null vectors.)
A Cartan spinor --- relative to such a decomposition! ---
is an element of the Grassmann algebra $\bigvee \3V_+$.
The spinor space $\3S$ then
has dimension
\BEq
\Dim \3S = 2^{{\big \lceil} {D-1\over 2}{\big \rceil}}
\EEq
where $\Dim \3V=D=2\Dim\3V_+$ and ${{\big \lceil} {N}{\big \rceil}}$
is the greatest integer $n\le N$.
Quantum theory interprets $\3V_{\pm}$ as mode spaces for a
fermion and
the Grassmann algebra $\bigvee \3V_+$ as
the algebra of
an aggregate of such fermions.
Thus Cartan spinors describe fermionic quantum aggregates,
not elementary individuals.

\subsection{Spinors describe aggregates}

The fact that a vector $v$ can be expressed
through a spinor $\psi$ bilinearly as $v^\mu=\psi^{\dag} \sigma^\mu \psi$ is
sometimes cited to indicate
that the vector is less elementary.
This is a categorical error.
The relation $v^\mu=\psi^{\dag} \sigma^\mu \psi$ indicates that a spinor,
like an aggregate of
vectors, carries enough information to define a vector.
But a vector cannot define a spinor.
Therefore the vectorial object is a subobject of the spinorial object,
not conversely.

Some have called Dirac's spinor space a ``square root of
space-time,'' because
$v^\mu$ is  quadratic in $\psi^\alpha$.
More accurately, a spinor space $\3S$
is the square root of the Clifford algebra $\3C=2^{\4X}\cong \3S\ox\3S\adj$
over space-time; or half thereof, after reduction:
\BEq
\3S=\sqrt{2}^{\3X}.
\EEq
$\Dim \3S$ is never less than $\Dim \3X$,
and is exponentially greater for high dimension.

Since plexors describe aggregates,
plexor theory can be a quantum substitute for set theory.
We apply it to quantum dynamics next.

\section{Clifford dynamics}

A {\em dynamics} too can  be defined
by a permutation.
First we  give a fixed set $\3V$ of elementary processes
that the system under study can undergo.
Then we give a permutation
\BEq
D: \3V\to \3V.
\EEq
For any elementary process $s\in \3V$, we interpret $Ds$ as the
immediate successor
of $s$ in the dynamical development defined by $D$.
We write the group of such permutations of $\3V$ as $\0S(\3V)$.

It is customary to avoid messy boundary questions
by imagining experiments that have gone on forever and will continue forever.
We shall use periodic boundary conditions.
Now the
experimental space-time region ultimately closes on itself,
outside the interesting part of the experiment.
This is no more fantastic than the infinite domain
and it permits us to work in a finite-dimensional
algebra.

A permutation $D$ partitions its domain $\3V$
into a congruence of orbits.
These are the threads that tie the dynamical elements together.
They do not intersect.
The intersecting geodesics
and the light-cones of space-time must
arise when we project
orbits from 8 dimensions down to
the 4 dimensions of space-time.
They arise from and describe a quantum entanglement
that occurs in the dynamical development.

Then a possible quantum concept of a dynamics mode $D$
is an operator $D$ in the algebra of a representation
of $\0S(\3V)$; for example in the algebra of the
covering group $\wt{\0S}(\3V)$.
Since $\0S_N$ is not simple in general,
and distinguishes a basis in $\SO_N$,
we do not found our theory on this concept.
Instead we imbed $\0S_N$ in the simple group $\SO_N$
as the axis-permuting elements,
and represent $\SO_N$.
We still interpret the operator $D$
as defining a dynamical succession.

Everything we know about a system
is in the record of our dynamical operations on the system.
A good language for quantum dynamics
is then a language of great expressive power.

As syntax for the dynamics language,
abstract $\dag$ algebra is not sufficient.
It deals only with how to combine
actions in series and parallel,
by multiplication and addition,
and how to reverse their internal chronological order,
by forming the adjoint.
It omits space-time fine structure,
which is supplied in standard quantum theory
by classical constructions prior
to the definition of the algebra.
How to express space-time concepts within the algebra
is part of the problem
of marrying quantum theory with gravity theory.
We use a high-dimensional
Clifford algebra to express quantum space-time.

The
statistics and the dynamics roads to the permutation group
join
when we postulate that ordinary dynamical processes
are aggregates too,
namely of elementary dynamical processes.
In the standard field theory the general process
is composed of operations that go on everywhere
in the system all the time,
described by the Hamiltonian or Lagrangian density.
Space-time is how the parts of the dynamics
are
interconnected.

Any quantum-dynamical theory must give the statistics
of its elementary quantum-dynamical processes.
In standard physics they
are tagged with space-time coordinates,
so they are distinguishable
and implicitly obey Maxwell-Boltzman statistics.

\subsection{Plexic dynamics}
\label{sec:SD}

To learn the structure
of the dynamical process,
we dissect it into
its atomic
constituents
and reassemble it out of them.
Evidently these elements of dynamics
must still have the nature
of dynamical processes themselves.
The dynamics is built out of elementary dynamical
actions $\chi$,
represented by arrows joining states.
The simplest classical
concept of dynamics is
a topological dynamics $\0D$.
This is usually presented as
a map $D:\0S\to \0S$ of a set of states $\0S$ into itself.
Instead we  deal only with dynamical actions
and not with states.
We define the dynamics as a mapping $D$
sending each arrow $\chi$
to its dynamical successor $D\chi$.

A theory that puts
states of being prior
to modes of action is called ontic,
the reverse {\em praxic}.
Praxism is an acute case of the
pragmatism
of Charles Peirce
and William James
and
the operationalism
of Einstein and Heisenberg.
Here we dissect dynamical operations
into micro-operations
   as James proposed to analyze experience
into micro-experiences.

This praxic concept of classical dynamics
introduces our two principals:
a set $X$ of atomic actions ${\chi\in X}$,
and a semigroup $\0S(X)$ of possible dynamical developments
\BEq
\label{eq:DYNAMICS}
D: X\to X.
\EEq
Dynamical developments are 1-1 mappings or permutations of $X\to X$.

To formulate a quantum dynamics
we ``quantize'' the structure
(\ref{eq:DYNAMICS}).
That is, we
replace classical variables described by sample spaces with
corresponding quantum variables
described by full matrix algebras.
Our prime variable is not space-time,
as Einstein proposed,
but the dynamical law.
Anandan (1999) has proposed (like Newton)
that dynamical law is variable and Smolin (1992, 1997)
that it evolves.
We sharpen this assumption
and take the dynamical alw as the only independent variable,
on which all others depend.

The individual elementary quantum process
making up the dynamics
we call the chronon $\chi$.
Its mode space $\3X$ and algebra $\End (\3X)$
replace the set $X$
of atomic actions.
Its operator algebra
$\3A({\3X}):=\End \3X$
replaces and synthesizes
the commutative Boolean algebra
of $X$
and the arrow semigroup of ordered pairs $X\x X$.
The nearest classical analogue
of a chronon
is not a space-time point,
which has no natural dynamical successor,
but a tangent or cotangent vector $(x, v)$
or $(x, p)$,
which does.
These form an 8-dimensional manifold,
not a 4-dimensional one.

To describe an aggregate
of chronons we need a statistics for the
chronon.

Neither Fermi,  Bose, nor Maxwell statistics will do.
A dynamics is a permutation.
A Fermi aggregate,
like a classical set,
is invariant under any permutation
of its elements.
It cannot
represent its dynamics by its permutations.
Nor can a Bose aggregate.

And Maxwell statistics
are reducible.

Evidently chronons,
to be permuted effectively,
must be distinguishable,
like classical space-time points,
which are implicitly supposed to have
Maxwell statistics.

In nature the ambient dynamics has modes
with spin 1/2.

The simplest statistics that supports
2-valued representations of $\0S_N$ is the
{\em Clifford statistics}\/.
The operator
algebra of this aggregate
is
a Clifford algebra
$\3C=\Cliff(\3V\0I)$
generated by individual unit modes $i_a$
obeying (\ref{eq:CLIFF})
The difference of two units
$C(ab) = i_a - i_b$
represents their swap $(ab)$.
We identify the individual mode space $\3V\0I$
with the first-degree subspace $\3C_1\subset \3C$.
Here there is no doubt
that the spinor represents an aggregate,
namely
the aggregate that the permutations permute.

Schur (1911) and Nayak \& Wilczek (1996)
use complex coefficients
throughout.
They represent some swaps by sums $i_a+i_b$
and others by differences $i_a -i_b$, depending on an arbitrary choice
of $N-1$ generating swaps.
It is not possible to represent all swaps $(ab)$ by sums $i_a+i_b$.
But their
representation is isomorphic to the
the simplex representation
(\ref{eq:SIMPLEX}),
of all swaps by differences.

Choosing Clifford statistics for chronons expresses
the distinguishability
of events
and the existence of spin 1/2.
The grade of a Clifford element gives the
minimal number of swaps or chronons in its factorization,
corresponding roughly to classical phase-space volume.

In the quantum theory of a variable dynamics D,
we distinguish between the dynamics D
and some dynamics operator $D$
that describes D maximally,
just as we distinguish between a hydrogen
atom H
and a mode vector ${\psi}$ maximally describing H.
A Hamiltonian is a kind of dynamics operator of the continuum limit.

Standard quantum theory uses modes in a
complex $\dag$ space $\3V$
whose $\dag$ defines
a non-singular sesquilinear form
$\psi\dag\phi$.
Gauge invariance requires
that the gauge generators be antihermitian,
and the gauge group structure
requires that some of them be nilpotent.
Only in an indefinite sesquilinear space
can a nilpotent other than the trivial 0
be antihermitian.
Therefore $\dag$ is usually indefinite,
and the quantum theories that work in Hilbert spaces,
with their definite $*$,
are not sufficiently relativistic for physics.

\subsection{Field theory under the microscope}

Clifford statistics also resolves a question
has beset quantum space-time physics from its inception.
What is the algebra of quantum fields on
a quantum space-time?
When we first asked this question
[Finkelstein (1969)] we imagined
that the q bits of the space-time
quantum computer all commuted,
and had serious difficulties with this question.
Now that they all anti-commute it answers itself.

First the problem. In classical physics, the
field fiber $\3F$ of field values
and the Minkowski manifold $\3M$
of space-time points
are combined (at least locally) by exponentiation
into a space
\BEq
\Phi=\3F^{\4M}
\EEq
of fields, each field $f\in \3F^{\4M}$ being a function $f: \3M\to\3F$.

Question: How do we define the exponential $F^M$
when the classical spaces $\3F$ and $\3M$
have been replaced by operator algebras describing
quantum field and space-time entities?

Answer: Take the logarithms of the algebras.
This reduces the computation of the exponential
to the computation of products,
an already solved problem.

We expand on this answer a bit.

One's first guess for the algebra $F^M$
is apt to be
the algebra of linear morphisms $M\to F$,
but this reduces merely to the tensor product
$M\adj\ox F$ when $M$ and $F$ are algebras,
and represents merely a pair of one $M$ quantum and one $F$ quantum.
This is to be expected, since
$M$ and $F$ describe individuals, not aggregates,
and a mapping from one individual to another
is merely one ordered pair.

For a better answer,
one must express algebraically
the fact that $\3M$
is a plenum, not a point.
All the points of $\3M$ are actual,
not mutually exclusive possibilities.

This is just the case for the Clifford statistics,
which permutes entities that are all present at once.
We designate a real free Clifford algebra $\3C$
over a quadratic space $\3X$ with endomorphisms algebra $\3A=\End(\3X)$ by
\BEq
\3C={\22}^{\4X}=\sqrt{\22}^{\4A}=\Cliff \3X\/.
\EEq
This makes $\3A$ a logarithm of $\3C$
and tells us how to define any quantum exponential of $\3C$:
\BEq
\label{eq:FIELD}
\3C^{\4M}:= (\sqrt{2}^{\4A})^{\4M} := (\sqrt{2})^{\4A\ox \4M}
=\Cliff(\3X\ox \3Y)
\EEq
where $\3M=\End (\3Y)$.

It is not hard to see
that the observed field algebras
do have logarithm algebras,
using the Chevalley (1954) representation of spinors within their
Clifford algebra.

{\em Definitions}\/:
An {\em octad} is a squad of 8 cliffordons with neutral quadratic form.
An octad space  is a real neutral quadratic space
\BEq
\28=\24\oplus\overline{\24}
\EEq
of 8 dimensions,
where the overline indicates a reversal of
metric, $\dag\to -\dag$.
An {\em octon} is a hypothetical quantum whose mode space is $\28$\/.
An {\em octadic space}
is a real neutral quadratic space
whose dimension is a multiple of 8:
The general octadic space is
\BEq
\3O=\28  \oplus \dots \oplus \28=N \28
\EEq
with $N>0$ terms.

\subsection{Examples}

The tangent-cotangent space to Minkowskian space-time $\3M$ is the octad space
   $\28=\1M\oplus\overline{ \1M\adj}$,
where $\1M:= \21\oplus\overline{\23}$ is the Minkowski tangent space.
The irreducible spinor spaces of $\28$ are again
octad spaces (Chevalley triality).

The Clifford algebra of an octadic space,
with its neutral quadratic form,
is algebra-isomorphic to the Clifford
algebra of a space of the same dimension
with a definite
quadratic form.

\paragraph{Octad lemma} An octadic chronon algebra
${\22}^{N\bf 8}$
factors as a Maxwell-Boltzmann ensemble
of $N$ octads each with algebra ${\22}^{\bf 8}$:
\BEq
\label{eq:OCTAD}
\22^{N\bf 8}=\22^{\bf 8} \ox \dots \ox \22^{\bf 8} \quad(N \mbox{ terms})
\EEq

In the limit $N\to\infty$,
this Maxwell-Boltzmann ensemble includes a
Bose-Einstein aggregate of octads.

It is easy to see that this Bose-Einstein aggregate admits condensation into
an 8-dimensional symplectic
manifold isomorphic to the tangent bundle to space-time.
A field of operators on space-time is
a similar condensation of a squad of octads
as $N\to \infty$, $\tav\to 0$.

This is a great simplification.
The group of a bundle is never simple;
the base couples to the fiber
without reverse coupling.
In Galilean relativity
the base was time,
while in field theory
the base is space-time,
but the illness is the same,
and the cure too:
relativization.
In (\ref{eq:FIELD}) the field and the space-time
are unified in the simple space-time-field
entity $\3S$.
When we first attempted to
express field theory in terms of q bits or chronons
[Finkelstein (1969)]
we imagined an absolute split between field fiber
and space-time base.
Now the field/space-time split appears to be
a factorization of a field-space-time unity $\3S$.
It is as relative
as the factorization of space-time into space/time.

\section{Quantification operators}

Each of the usual statistics,
Fermi, Bose and Maxwell,
has an operator-valued form
$Q\adj$ and dual form $Q$
that defines how
infinitesimal actions on the individual
can act on
the aggregate.

In each statistics the individual I
has a mode vector space $\3V(\0I)$ and operator algebra $\3A(\0I)$.
The aggregate has a mode vector space $\3V(\0S)$
and operator algebra $\3A(\0S)$.
Let $d\3A$ be the infinitesimal Lie algebra of $\3A$ with Lie product
$[\al,\be]:=\al\be-\be\al$.
The quantification operator
$Q\adj$ is
a linear morphism
\BEq
Q\adj: \3V(\0I) \to
\3A(\0S), \; \psi \mapsto Q\adj \psi,
\EEq
transforming a mode vector (or a ket) $\psi\in \3V(\0I)$ for the individual
to an operator $Q\adj \psi\in \3A(\0S)$ for the aggregate.
The operators $\psi\adj Q$
generate $\3A(\0S)$ $\dag$-algebraically.
The mapping
\BEq
\label{eq:LIE}
   \bra {\dots}\ket: d\3A(\0I)\to
d\3A(\0S),  w\mapsto \wh{w}=Q\adj w Q
\EEq
is a Lie-algebra homomorphism.

We call the $Q$ with these properties, when it exists,
the {\em quantification operator}
of the statistics.
If ${w}$ represents
an additive quantity
or infinitesimal transformation
of the individual,
we call $Q\adj {w} Q$
the {\em quantified} ${w}$ for the quantified system.
The quantification operators of Fermi, Bose and Maxwell statistics
map individual mode vectors $\psi$ into annihilation operators $Q\adj \psi$\/,
and individual quantities $q$ into additive total quantities $Q\adj q Q$\/.
Clifford statistics
also has a quantification operator $Q\adj$,
which maps
mode vectors into
swaps instead of creation operators.

The Clifford quantification operator $Q_0$ obeys the Clifford law
\BEq
(\forall v\in \3V)\quad (Q_0\adj v)^2 = ||v||
\EEq
We chose the sign in (\ref{eq:CLIFF}) so that the mapping
(\ref{eq:LIE})
is a Lie algebra homomorphism,
preserving the commutation relations
of the individual
within those of the aggregate,
as for Fermi and Bose statistics.
This is just the familiar fact that the
commutators $L_{ab}=i_{[ab]}$
generate a representation of the orthogonal group.

We write the quantification operators for
Clifford, Fermi,  and Bose statistics
as $Q_0, Q_1$, and $Q_2$.
The numerical subscripts count
the independent imaginaries
in the coefficient field $\1R, \1C$, and  $\1H$
of the classical group
of the statistics.
We write $Q_{\6M}$ for the Maxwell-Boltzmann quantifiation operator.
We call the most important aggregates (Maxwell) {\em sequences}\/,
(Bose) {\em sibs}\/, (Fermi) {\em sets} and (Clifford) {\em squads}
for brevity.
We call the Clifford composite a s{\em qua}d
to remind us that it is an essentially {\em qua}ntum structure.
There are classical sets and
classical sibs, but no classical squads.
Clifford statistics,
like anyonic and other multivalued statistics,
involves quantum
superposition more deeply than the
single-valued composites such
as the sequence, sib, or set.

For example if ${L}$ is
a component of individual angular momentum,
then for Fermi quantification $Q=Q_1$ and Bose quantification $Q=Q_2$,
$Q^\adj {L} Q$ is the total angular momentum of the aggregate.
These quantifications {\em totalize}
the operator on which they act.

The Maxwell quantification operator $Q_{\mbox{\rmss M}}$ does not totalize.
The quantified operator $Q_{\mbox{\rmss M}}\adj {L} Q_{\mbox{\rmss M}}$
represents the ${L}$
of only the last individual in the sequence,
not the total ${\omega}$.
Totals have somewhat more complicated expressions
in Maxwell statistics.

Clifford quantification, like Fermi and Bose, totalizes.

Clifford statistics
relates a quadratic space $\3V$,
its endomorphism algebra $\3A$,
its Clifford algebra $\3C$,
and its spinor space $\3S$,
by the commutative diagram
\BEq
\begin{array}{rlcrl}
\hphantom{{}_{ \Sq }}{\3V}& \stackrel{\End}{\to} &
\hphantom{{}_{ \Sq }}\3A\cr
\mbox{\scriptsize $ \Sq $}\kern-3pt\dar&
{\kern4pt}\stackrel{}{\searrow}{\kern-6pt} \stackrel{\Cliff}{}
&\mbox{\scriptsize$ \Sq $}\kern-3pt\dar\cr
\hphantom{{}_{\Sq }}\3S &\stackrel{\End}{\to}
&\hphantom{{}_{\Sq }}\3C
\end{array}
\EEq

When we apply Clifford statistics
to dynamics,
$\3V$ is the mode space $\3X$
for a chronon.
The composite system described by a spinor of $\3S$
consists
of all the chronons transpiring
in the experimental space-time region.
The algebra $\3A$ consists of endomorphisms of $\3V$.
The Clifford algebra $\3C$
consists of descriptions $D$ of the global dynamics
of the squad.

The real Clifford algebra $\3C=\Cliff(\3V,\1R)$
of Clifford statistics
is the endomorphism algebra
of an underlying spinor module $\3S$
over one of the five rings $\1R, \1C, \1H, 2\1R, 2\1H$,
depending on the dimension and signature of $\3V$\/
according to
the spinorial chessboard [Budinich \& Trautman (1988)].

Clifford statistics can readily simulate Bose and Fermi
statistics with pseudo-bosons and pseudo-fermions.
In (\ref{eq:OCTAD}), Clifford statistics
exactly simulates
an aggregate of octads obeying mutual Maxwell-Boltzmann statistics
and internal Clifford statistics at the same time.

The most striking difference
between the Clifford statistics
and Fermi
statistics
cannot be read from their algebras.
Orthogonal ${\psi}$'s anticommute
in both statistics,
and every fermionic algebra is
algebra-isomorphic to a neutral Clifford algebra,
but fermions are identical and cliffordons
are not.
An algebra does not define its interpretation.
The exchange of two fermions is represented
by factor exchange, one stipulates, and hence by $-1$, which is
projectively equivalent to the identity.
The exchange of  cliffordons 1 and 2, however,  is
represented by ${i}_1-{i}_2$.
Cliffordon swaps, far from being trivial, scalar or central,
may generate
the entire aggregate action algebra.

\section{Chronon dynamics}
\label{sec:CHRONON}

We  hypothesize that
the dynamics of a
suitably isolated
physical system
is a squad of
elementary dynamical processes or chronons,
and that the
ambient vacuum breaks
its Clifford algebra $\3C$ down
into many
mutually commuting
local octadic Clifford algebras
as in (\ref{eq:OCTAD}).

We construct a simple finite-dimensional Clifford algebra
$\3C=\breve{\kern-4pt\3A}$
that approaches (or ``contracts to'')  the Minkowski manifold algebra,
the associative algebra  $\3A=\3A(x^{\mu}, \pd_{\mu})$
of coordinates $x$ and derivations $\pd$
of space-time differential geometry.
$\3A$ may be regarded as a generalization
of the Heisenberg algebra of $x$ and $p$  and
a variant of the Bose-Einstein algebra.
Expanding it
into a Clifford algebra
is mathematically akin to approximating bosonic fields with fermionic ones,
the process of
bosonization.

Since $\3A$ is infinite-dimensional,
the dimensions of $\3C$ and its orthogonal group
are huge, like the
number of phase-space cells in the experiment, which
is likely
$\gg 10^{20}$ for atomic experiments,
and approach infinity in the contraction to the continuum.

Chronons and the basic Clifford variables $i_a$
that represent them
are pre-local
in the extreme,
since they all anticommute.
Nevertheless they are the raw material of our universe, we propose.

This only apparently clashes with quasilocality,
due to (\ref{eq:OCTAD}).

\subsection{Localization}

We begin construction with an octadic chronon space $\28 N$
and its Clifford algebra $\3C$.

We decompose $\3C$ into
$N$ mutually commuting octad algebras of independent local
variables $\gamma_{\mu}(n)$ obeying local commutation relations
\BEq
\{\gamma_{\nu}(n), \gamma_{\mu}(m)\} = t_{\nu\mu}\delta_{nm}
\EEq

We define the Lorentz algebra and group of such a Clifford algebra
by the usual expression
for the angular momentum
of a spin aggregate,
\BEq
\label{eq:LEXP}
\breve{L}_{\nu\mu} :={1\over 2}\sum_{n=1}^N
\sum_{\be=0}^1\gamma_{\nu\mu}(n,\be)
\EEq

We assume that  $p_{\mu}$ and $x^\mu$ for the $N$ squads
are,
like $L_{\mu\nu}$,
   additively composed of terms from each squad,
corresponding to how the displacement $\Delta x =\int_C dx(\tau)$
along a curve $C:
x=x(\tau)$ is an integral  over $C$
of a contribution from each differential element $dx(\tau)$:
\BEq
\Delta x^\mu = \sum_{n,\be} \delta x^\mu(n,\be)\/,\quad
\Delta p_\mu = {1\over 2N}\sum_{n,\be} \delta p_\mu(n,\be)\/.
\EEq

Here we use $\hbar = c = \tav= 1$.

Suppose
each tetrad has  contributions
\BEq
\delta x^\mu = 2^{-1/2} \gamma^\mu,\quad
\delta p_\mu =  2^{-1/2} \gamma^\uar \gamma_\mu,\quad
\de i = \gamma^{\uar}.
\EEq
The unit of $x$ is $\tav$ and the unit of $p$ is
$\hbar/\tav$\/, while $i$ is dimensionless.

Then for each tetrad
\BEq
\label{eq:XHCR}
\begin{array}{ccccccc}
[\delta x^{\nu}, \delta p_{\mu}]
   & = &\delta i \;\delta_{\nu\mu}\/,\cr
[\de i, \delta x^\mu]  & = & +2 \delta p^{\mu}\/,\cr
[\de i, \delta p_\mu] & = & -2\delta x_{\mu}\/,
\end{array}
\EEq

The first of equations (\ref{eq:XHCR})
makes $\gamma^\uar$ the expansion of
Heisenberg's
$i$, much as Hestenes (1966) proposed.
Presumably this builds in a violation of parity.

The second and third tell us that
the expanded $i$ generates the symplectic symmetry between
$x$ and $p$,
as Segal (1951) proposed.
On the chronon scale, this violates Heisenberg's commutation
relations seriously.
In the standard quantum theory $i$ is central.

We  recover the Heisenberg commutation relations as a contraction of
the Clifford
relations by summation and a subsequent correlation:
\BEqA
\label{eq:XHOP}
\breve{x}^{\mu} &=& \sum_{n,\be} \delta x^\mu(n,\be), \cr
\breve{p}_{\mu} &= &{1\over 2N}\sum_{n,\be} \delta p_\mu (n,\be),\cr
\breve{i}&=&\sum_{n,\be}\de i (n,\be)
\EEqA

Here and in what follows,
the breve on a variable,
like $\breve{i}$ above,
is a semantic annotation rather than a syntactic one.
It declares that the accented variable is
an expansion of the standard unaccented one,
and reduces to it upon contraction.
It thus tells us how to measure the variable
in the contracted domain of experience.
If one set of physical concepts successfully covers both domains,
as in quantum theory and relativity, for example,
we drop such labels.

Feynman (1971) proposed
that the space-time coordinate-difference operator
is the sum of many mutually commuting tetrads of Dirac vectors:
\BEq
\label{eq:FEYNMAN}
\Delta x^\mu = \mbox{Const}\sum \gamma^\mu(n)
\EEq
as a quantum form of the proper-time Heisenberg-Dirac
equation $dx^\mu/d\tau =\gamma^\mu(\tau)$.
Feynman's proposal returns in (\ref{eq:XHOP}) as a result of the octad lemma.
Each of his $\gamma$'s represents one chronon $i^a$ in a large octadic squad.

\subsection{Correlation}

We now have far too many $i$'s: one for every tetrad.
In the standard physics
there is only one imaginary $i$ and one Clifford vector $\gamma^{\mu}$
for the whole system.
We correlate all the $\de i$ so that they are effectively one $i$,
and all the tetrads $\gamma^\mu(n)$
so that they are effectively one tetrad $\gamma^\mu$.
We suppose that the vacuum Bose-Einstein condensation establishes
this correlation.
The local departure of $\breve{i}$ from its global mean $i$
is presumably a Higgs operator.

To be sure, the tetrads of the tetrad lemma obey Maxwell statistics, not Bose.
But the mode space of Maxwell statistics is the direct sum of the
spaces of all the
tensorial statistics,
including Fermi and Bose and parastatistics.
Any Bose projection is a fortiori a Maxwell projection.

We require a projection operator  $P\in\3C$
expressing this alleged Bose-Einstein correlation,
projecting onto the symmetric subspace.

We have constructed this symmetrizer elsewhere [Finkelstein (2000)].
The symmetric subspace of $\3S$ is $P\3S$,
the  $P$-image of $\3S$.
The restrictions of the $\breve{i}(n)$ to $P\3S$ are all equal to
$P\breve{i}P$,
the restriction of $\breve{i}$ to $\3S$,
with $\breve{i}$ given by (\ref{eq:XHOP}).

It is then straightforward to see that this restriction of
$\breve{i}$ is a square root of
the restriction of $-1$ to
$P$:
\BEq
(P\breve{i}P)^2 = -P
\EEq

It remains to be seen whether the variable $P\breve{i}P$ is
sufficiently close to being
central in processes close to the vacuum.
If so then it can pass for the physical $i$ of quantum
mechanics in such processes.

The Clifford elements
\BEq
L_{\mu\nu} = \sum_{\beta,\tau} [i_{\mu,\beta,\tau}, i_{\nu, \beta, \tau}]
\EEq
are infinitesimal generators of the connected Lorentz group $\3L$
of this model.
The operators $x$ and $p$ constructed by (\ref{eq:XHOP}) are covariant under
the connected Lorentz group $\3L$\/.

\section{Relativistic dynamics operator}

To recover Maxwell statistics in the classical space-time limit
we were forced to order the octads
with a ``proper time'' index
$\tau$.
This permits us to formulate an ``octad-cycling'' dynamics operator
$D_o$ that advances
$\tau$ by unity and so is a scalar invariant,
unlike the Hamiltonian,
which increases $t$ and is one component of a vector.

A covariant development in proper time is generated by a {\em rest
mass operator} just as a
coordinate-time development is generated by a Hamiltonian energy operator.
A covariant proper-time
dynamics with a second-order mass operator is used by Schwinger
(1949) among others.
The present theory is a kind of ``quantum square-root'' of such theories.

$D_o$ is then an ordered product of 8 disjoint cycles of length $N$
according to Clifford
statistics.
Defining $\Delta i_{\mu, \beta, \tau}:= (i_{\mu, \beta, \tau +1} -
i_{\mu, \beta, \tau})$,
we write this basic dynamics operator as
\BEq
\label{eq:DYN}
D_o=\prod_{\beta}^{\lar} \prod_{\mu}^{\lar} \prod_{\tau}^{\lar}
\Delta i_{\mu, \beta, \tau}.
\EEq
In the $\tau$-ordering, $\Delta i_{\mu, \beta, \tau}$ is to be
treated as a whole with
one index $\tau$,
not divided between $\tau$ and $\tau+1$.
Each cycle of (\ref{eq:DYN})
swaps the $\mu, \beta$ chronon of octad $\tau$
with the  $\mu, \beta$ chronon of octad $\tau+1$
according to the orthonormal statistics (\ref{eq:DIFF}).

Now  two apparently
separate conceptual streams, special relativity
and combinatorics, merge almost unexpectedly:

{\bf The octadic chronon dynamics $D_o$ is Lorentz invariant.}

The proof is straightforward. It rests on the familiar
fact that the Dirac top gamma $\gamma^{\uar}=$``$\gamma^5$''
is invariant under the connected Lorentz group.
This confluence gives us
renewed hope for a chronon dynamics.

$D_o$ has several easily constructed brothers that are also Lorentz-invariant
and also shift tetrads or octads forward in $\tau$ by small steps.

$D_o$ is still unsatisfactory for many reasons.
Above all, it does not define the Minkowskian
metrical structure of space-time.
In a simple theory,
Poincar\'e invariance of the ambient dynamics operator $D$ is not enough.
$D$ must also define the space-time metric.

Physically, we must know the momentum of a moving particle
to predict
its next position sharply.
This means a coupling among the 8 components of
space-time-energy-momentum.
The dynamics $D_0$ is an uncoupled development.
Lorentz-invariant couplings are now under study.

\section{Simplifying the standard model}

The simplification of the space-time structure of the standard model
that we have performed so far
suggest that we can simplify its internal structure as well
by dissolving the separation between field variables and space-time variables.
This was the goal of our
earlier Fermi quantizations of space-time,
and it comes closer in the present
Clifford quantization.

The
standard theory needs internal variables
to describe hypercharge,
isospin, color, and family because it
assumes that
the
immediate neighborhood of any event
is exactly Minkowskian.
{\em At any one point
all gauge vectors
can be transformed to zero
by a gauge transformation.}

In the case of gravity this assumption
of the standard model
specializes
to Einstein's equivalence principle.
We call this italicized proposition
the {\em generalized equivalence principle}
when we intend to include gravity among the gauge fields.

We have supposed that the field variables
actually describe finite quasilocal defects of size $\tav$
in the vacuum condensate.
In the continuum limit $\tav\to 0$,
these vanish,
and surrogate variables have to be invented.
Variables that describe
the condensate in maximal quantum detail,
with $\tav\ne 0$,
should suffice to describe
its defects.

This  approach to simplification
avoids the compactification problem
that plagues theories of the Kaluza kind.
It replaces mysteriously small extra dimensions
by physically small neighborhoods.
The same chronon variables $i^a$
that combine to form the external $x$'s and $\pd$'s,
also combine into the internal $\gamma$'s and $\tau$'s,
the fields, and the Lagrangian,
a contraction of the dynamics operator.
In the condensate many of these degrees of freedom
are frozen.
A high-energy interaction thaws and excites
some of them.
We require that the same $i$'s give us
the external modes
when they sing in unison
and the internal modes when they sing
polyphonically.

\section{Discussion}

The revolutions of the past hundred years of physics have simplified
certain algebras.
Several non-simple algebras still wait to be simplified,
notably
the Heisenberg algebra of quantum theory,
the algebra of the field bundle,
and the algebra of dynamics.
These couple to each other
so that the next revolution must likely simplify them all at once.
The result will be a non-local quantum theory.

We have given one obvious simplification
of these algebras,
postulating a dynamics operator $D$
that contracts to the standard Hamiltonian and Lagrangian
in a suitable limit
but consists of many
elementary quantum actions,
chronons.
The key unifying element
is the Clifford statistics for
the chronon.

This leads us to a quantum correspondent
for the standard gauge principle:
Remote comparisons are effected by
a succession of quantum swaps.
It also suggests a host of
relativistic candidates
for the dynamics operator
$D$\/.

\section{Acknowledgments}

This work was stimulated by discussions with
Dennis Marks,
   Mohsen Shiri,
Frank D. (Tony) Smith,
and
Zhong Tang, and others in the Quantum Relativity Workshop
at Georgia Tech,
and with
J. Anandan,
Giuseppe
Castagnoli,
Lee Smolin,
Raphael Sorkin,
and  Frank Wilczek.
It was supported by
the Institute for Scientific Interchange,
the Elsag-Bailey Corporation,
the M. and H. Ferst Foundation,
and the University System of Georgia.

\end{document}